\documentclass[twocolumn,showpacs,preprintnumbers,pra,amsmath,amssymb]{revtex4}

\usepackage{graphicx}% Include figure files
\usepackage{bm}% bold math
\begin{document}
\title{$\psi$-Epistemic Models and Bell Theorem}

\author{Won-Young Hwang\footnote{Email: wyhwang@jnu.ac.kr}}

\affiliation{Department of Physics Education, Chonnam National University, Gwangju 61186, Republic of Korea
}
%%%%%%%%%%%%%%%%%%%%%%%%%%%%%%%%%%%%%%%
\begin{abstract}
We consider a question in what condition a mixed state which can be decomposed in different ways cannot be described by a single set of hidden variables. The condition is closely related with Bell theorem.
\end{abstract}
%%%%%%%%%%%%%%%%%%%%%%%%%%%%%%%%%%%%%%%
\maketitle
%%%%%%%%%%%%%%%%%%%%%%%%%%%%%%%%%%%%%%%
\section{Introduction}
%%%%%%%%%%%%%%%%%%%%%%%%%%%%%%%%%%%%%%%

Recently Spekkens studied possibility of purely epistemic models for quantum mechanics \cite{Spe07}. Because highly non-orthogonal pure states give almost the same outcomes for measurements, it appears to be reasonable that sets of hidden variables corresponding to the two states overlap. However, as shown by Pusey-Barrett-Rudoph (PBR) \cite{Pus12} later, no such overlap is allowed as long as state-independence, a reasonable assumption, is satisfied.

It is well known that a mixed state can be expressed in many different ensemble decompositions \cite{Nie00}. In toy model in Ref. \cite{Spe07}, sets of hidden variables for mixtures with different decompositions are the same as long as the mixtures correspond to the same density operator. For example, equal sum of hidden variable sets for states $|0\rangle$ and $|1\rangle$ is the same as the sum of hidden variable sets for states $|+\rangle$ and $|-\rangle$. Here $|0\rangle$ and $|1\rangle$ are two orthonormal states of a quantum bit (qubit) and $|\pm\rangle= (1/\sqrt{2})(|0\rangle \pm|1\rangle)$. This property appears to be reasonable. However, this is not the case in general \cite{Lei11} because each state is described by disjoint set of hidden variables according to PBR theorem.

However, the property can still hold for limited number of states and measurements, as in the toy model. This is not in contradiction with PBR theorem because implication of the theorem is that hidden variable sets must be disjoint in order to be fully consistent with quantum mechanics. Now a question is under what conditions the property is violated. In this paper, we will show that for certain three projection measurements and associated pure states, the property is violated. We will show how the violation is closely related with Bell theorem.

%%%%%%%%%%%%%%%%%%%%%%%%%%%%%%%%%%%%%%%
\section{Results}
%%%%%%%%%%%%%%%%%%%%%%%%%%%%%%%%%%%%%%%
Let us consider $Z$, $X$, $(Z+X)/\sqrt{2}$ measurements which are composed of $|0\rangle \langle 0|$ and $|1\rangle \langle 1|$, $|+\rangle \langle +|$ and $|-\rangle \langle -|$, $|\pi/4\rangle \langle \pi/4|$ and $|5\pi/4\rangle \langle 5\pi/4|$, respectively.
Here $|\theta \rangle= \cos(\theta/2) |0\rangle+ \sin(\theta/2)|0\rangle $ and Bloch vector of  $|\pi/4\rangle$ is in between two others of $|0\rangle$ and $|+\rangle$. We also consider six pure states associated with the measurements.
Each state can be measured by one of the measurements. Let us consider hidden variable models which are compatible with all outcomes of the measurements on the states. Now let us consider equal mixture of $|0\rangle$ and $|1\rangle$, that of $|+\rangle$ and $|-\rangle$, that of $|\pi/4\rangle$ and $|5\pi/4\rangle$. Density operators of the three decompositions are the same. The question is, as discussed, whether set of hidden variables corresponding to the three decompositions can be the same as far as the three measurements are concerned.

%%%%%%%%%%%%%%%%%%%%%%%%%%%%%%%%%%%%
{\bf Proposition-1}:
%%%%%%%%%%%%%%%%%%%%%%%%%%%%%%%%%%%%
Hidden variable models compatible with outcomes of the measurements on the associated states cannot satisfy the property that set of hidden variables assigned to the three different decompositions are the same.

Let us show Proposition-1. Let us consider states $|\pi/4\rangle$ and $|5\pi/4\rangle$. Without loss of generality, we can suppose that space of hidden variables $\lambda$ is confined to between $0$ and $1$, namely $0\leq \lambda < 1$ and probability distribution of the hidden variable is constant. Also without loss of generality, sets $\{\lambda|0\leq \lambda < 1/2 \}$ and $\{\lambda|1/2 \leq \lambda < 1 \}$ can be assigned to states $|\pi/4\rangle$ and $|5\pi/4\rangle$, respectively. That is, for $Z+X$ measurement, hidden variables between $0$ and $1/2$ (between $1/2$ and $1$) gives an outcome $0$ (an outcome $1$).
Here for convenience we omit normalization for measurement.
Now let us consider hidden variable sets for states $|0\rangle$ and $|1\rangle$. Suppose that the property is satisfied. Then we can see that the same probability distribution of hidden variable within the same space $0\leq \lambda < 1$ should be assigned to equal mixture of the states $|0\rangle$ and $|1\rangle$. However, for $Z+X$ measurement the $|0\rangle$ state should give outcome $0$ with probability $\cos^2(\pi/8) \approx 85\%$. This means that $|0\rangle$ and $|\pi/4\rangle$ share around $85\%$ of hidden variables. That is, around $85\%$ of $|0\rangle$ state's hidden variable is composed of those between $0$ and $1/2$. Similarly, we can see that $|+\rangle$ and $|\pi/4\rangle$ also share around $85\%$ of hidden variables. Combining the two, we obtain that $|0\rangle$ and $|+\rangle$ states share around $70\%$ at least. This implies that when $Z$ measurement is performed on $|+\rangle$ state, probability to get outcome $0$ is around $70\%$ at least. This is in contradiction with quantum mechanical predictions. $\Box$

What we have discussed is closely related to the followings. Prepare a Bell state $|\varphi^+\rangle= (1/\sqrt{2})(|0\rangle_A |0\rangle_B +|1\rangle_A |1\rangle_B$). Here $A$ and $B$ denote Alice and Bob having each qubit. At each time, Alice and Bob independently and randomly choose one among $Z$, $X$, $Z+X$ measurements and perform the chosen one on each qubit.

%%%%%%%%%%%%%%%%%%%%%%%%%%%%%%%%%%%%
{\bf Proposition-2} (Bell theorem):
%%%%%%%%%%%%%%%%%%%%%%%%%%%%%%%%%%%%
No local hidden variable model can simulate outcomes by the measurements on the Bell state.

Proposition-2 follows from the fact that a Bell inequality violation is hidden. The outcomes of the measurements can violate a version of Bell inequality \cite{Sak85} $P(01|Z,Z+X)+ P(01|Z+X,X) \geq P(01|Z,X)$, because here $P(01|Z,Z+X)=P(01|Z+X,X) \approx 15\%$ and $P(01|Z,X) = 50\%$. Here $P(01|Z,Z+X)$, for example, denotes probability to get outcomes $0$ and $1$ when measurements $Z$ and $Z+X$ are done, respectively. $\Box$

Now let us discuss how the propositions are related.

%%%%%%%%%%%%%%%%%%%%%%%%%%%%%%%%%%%%
{\bf Proposition-3}:
%%%%%%%%%%%%%%%%%%%%%%%%%%%%%%%%%%%%
Proposition-1 is equivalent to Proposition-2.

Suppose that Proposition-1 is violated. Namely, let us suppose  without loss of generality that a hidden variable model with hidden variable $0\leq \lambda < 1$  with constant probability distribution can reproduce outcomes of the measurements on the mixtures. Then the hidden variable model can be immediately utilized by Charlie in between Alice and Bob to violate Proposition-2. Two  physical entities (e.g. particles) having the same hidden variable $\lambda$ are prepared according to the probability distribution, then one is sent to Alice and the other to Bob. We can easily see that this extended hidden variable model can simulate outcomes of the measurements in Proposition-2. For example, let us consider the case when Alice performs $Z+X$ measurement. Assume that $0\leq \lambda < 1/2$ ($1/2\leq \lambda < 1$). Then Alice gets an outcome $0$ (1). Here what Bob receives amounts to a state $|\pi/4\rangle$ ($|5\pi/4\rangle$). So outcomes can be reproduced for any one of the measurements chosen by Bob. Other cases can be similarly explained.

Reversely, suppose that Proposition-2 is violated. Then any (local) hidden variable simulating the measurements on the Bell state can be adopted to violate Proposition-1. Let us group set of the hidden variable according to the outcomes of the measurements. Let $\Lambda_{A_z A_{z+x} A_x}^{B_z B_{z+x} B_x}$ be a set of all hidden variables $\lambda$ which gives outcomes $A_z, A_{z+x}, A_x$ for Alice's $Z, Z+X, X$ measurements and outcomes $B_z, B_{z+x}, B_x$ for Bob's $Z, Z+X, X$ measurements, respectively. Here value of $ A_z,A_{z+x},A_x,B_z,B_{z+x},B_x$ can be either $0$ or $1$. For example, $\Lambda_{000}^{101}$ denotes set of all hidden variables giving outcomes $0,0,0$ for Alice's $Z, Z+X, X$ measurements and outcomes $1,0,1$ for Bob's $Z, Z+X, X$ measurements, respectively. Then let us consider dividing the set of hidden variables according to Bob's $Z+X$ measurement, for example. Let sum of $\Lambda_{A_z A_{z+x} A_x}^{B_z 0  B_x}$ for all $A_z, A_{z+x}, A_x, B_z, B_x$ values be denoted by $\Lambda_{\diamond \diamond \diamond}^{\diamond 0 \diamond}$. Now we can see that the hidden variable set  $\Lambda_{\diamond \diamond \diamond}^{\diamond 0 \diamond}$ properly works as the $|\pi/4\rangle$ state for the measurements by Alice. That is, the hidden variable set  $\Lambda_{\diamond \diamond \diamond}^{\diamond 0 \diamond}$ reproduces outcomes of the measurements by Alice. Other cases are similar. Hidden variable set corresponding to equal mixture of $|\pi/4\rangle$ and $|5\pi/4\rangle$ states is sum of $\Lambda_{\diamond \diamond \diamond}^{\diamond 0 \diamond}$ and $\Lambda_{\diamond \diamond \diamond}^{\diamond 1 \diamond}$, which is set of all hidden variable, $\Lambda_{\diamond \diamond \diamond}^{\diamond \diamond \diamond} \equiv \Lambda$. Hidden variable sets  corresponding to other two decompositions, namely equal mixture of $|0\rangle$ and $|1\rangle$ states, and that of $|+\rangle$ and $|-\rangle$ states, are all identical to $\Lambda$. So Proposition-1 is violated. $\Box$

%%%%%%%%%%%%%%%%%%%%%%%%%%%%%%%%%%%%%%%
\section{Discussion}
%%%%%%%%%%%%%%%%%%%%%%%%%%%%%%%%%%%%%%%
Combination of Proposition-1 and Proposition-3 amounts to another proof of Bell theorem. That is, supposing existence of local hidden variables reproducing the measurements on the Bell states implies existence of hidden variable models for a single qubit that violates Proposition-1.

In conclusion, we have shown that, in a specific but generic example, a mixed state which can be decomposed in many different ways cannot be described by a single set of hidden variables.  We showed how the fact is related with Bell theorem.
\section*{Acknowledgement}
This study was supported by Institute for Information and Communications Technology Promotion (IITP) grant funded by the Korea Government (MSIP) (No. R0190-16-2028, Practical and Secure Quantum Key Distribution).
%%%%%%%%%%%%%%%%%%%%%%%%%%%%%%%%%%%%%%%%%%%%%%%%%%%%%%%%%%%%
%%%%%%%%%%%%%%%%%%%%%%%%%%%%%%%%%%%%%%%%%%%%%%%%%%%%

\end{document}